\documentstyle[aps,prl,twocolumn]{revtex}
\begin{document}
\draft
\title{Mutually Penetrating Motion of Self-Organized 2D Patterns of Soliton-Like
Structures}
\author{K. Indireshkumar and A. L. Frenkel\cite{byline}}
\address{Department of Mathematics, University of Alabama, Tuscaloosa, Alabama\\
35487-0350}
\date{\today }
\maketitle

\begin{abstract}
Results of numerical simulations of a recently derived most general
dissipative-dispersive PDE describing evolution of a film flowing down an
inclined plane are presented. They indicate that a novel complex type
of spatiotemporal patterns can exist for strange attractors of
nonequilibrium systems. It is suggested that real-life experiments
satisfying the validity conditions of the theory are possible: the required
sufficiently viscous liquids are readily available.
\end{abstract}


\preprint{HEP/123-qed}

\narrowtext
The phenomenon of pattern formation in nonequilibrium dissipative systems is
currently a topic of active experimental and theoretical research (see e.g. 
\cite{ch93} for a recent progress review). Here we report our theoretical
studies and numerical simulations of a two-dimensional (2D) evolution PDE
approximating a flow down an inclined plane; it exhibits self-organization
of a remarkably complex spatiotemporal pattern which then persists indefinetly
in this dissipative-dispersive system. In certain cases discussed below,
such a pattern consists of two {\it subpatterns} of two-dimensionally (2D)
localized surface structures. One of these subpatterns is an essentially 1D
arrangement of larger-amplitude {\it bulges} on the film surface which are
nearly equidistantly aligned on (a number of) straight-line segments; those
are surrounded by smaller-amplitude {\it bumps}, which constitute the
second, lattice-like subpattern filling up essentially the entire flow
domain. Each of the two subpatterns moves as a whole; their velocities are
different, and every elementary structure (a bulge as well as a bump)
periodically collides with those of the other kind. In the collision of a
bump with a bulge (or with a pair of neighboring bulges), the two structures
pass through each other similar to the well-known 1D Korteweg-deVries
solitons, returning to their pre-collisional shapes and speeds after the
interaction.

Studies of wavy film flows on solid surfaces (the ``Kapitza problem'') have
a considerable history. However, the nonlinear dynamics of wavy films is far
from being fully understood (see e.g. \cite{ls95}; see \cite{fi96} and \cite
{cd96} for recent progress reviews). Fortunately, the nonlinear coupled-PDE
Navier- Stokes (NS) problem, additionally complicated with a free boundary,
can be reduced to simpler {\it approximate} descriptions of the wave
dynamics for certain domains of the parameter space. In the most favorable
cases, such a description reduces to a {\it single} partial differential
equation (PDE) governing the evolution of film thickness. Recently, we (see 
\cite{fi96}) have derived the most general evolution equation (EE) capable
of all-time-valid description of a wavy liquid film (of a constant density $%
\rho $, kinematic viscosity $\nu $, surface tension $\sigma ,$ and average
thickness $h_0$) flowing down an inclined plane. Its dimensionless form is 
\[
\eta _t+4\eta \eta _z+\frac 23\delta \eta _{zz}-\frac 23\cot \theta \eta
_{yy}+\frac 23 W\nabla ^4\eta 
\]
\begin{equation}
+2\nabla ^2\eta _z=0.  \label{e33}
\end{equation}
Here $\eta $ is the deviation of film thickness from its average value of $1$%
, $z$ and $y$ are the streamwise and spanwise coordinates, and $\nabla
^2=\partial ^2/\partial z^2+\partial ^2/\partial y^2$. We have defined $%
\delta \equiv (4R/5-\cot \theta )$, where $\theta $ is the angle of
inclination of the plane and $R=h_0U/\nu $ is the Reynolds number. Here $%
U=gh_0^2\sin \theta /(2\nu )$ where $g$ is the gravity acceleration, and in
Eq. (1) $W=\sigma /(2\rho \nu U)$ is the Weber number. All the dimensionless
variables are measured in units based on $h_0,$ $U$, and $\rho $. Equation
(1) describes the film evolution in a reference frame moving with the
velocity $2U$ in the streamwise direction.

In this paper, we limit ourselves to the case of $\theta =\pi /2,$ i.e.
flow down a vertical wall. Then, we can transform the EE to a ``canonical''
form which will contain only{\it \ one} control parameter---by rescaling $%
\eta =N\widetilde{\eta }$, $z=L\widetilde{z}$, $y=L\widetilde{y}$, and $t=T%
\widetilde{t}$, where $N=2R/(5W)$, $L=\sqrt{5W/(4R)}$, and $%
T=(5^{3/2}/16)(W/R)^{3/2}$. Dropping the tildes in the notations of
variables, the resulting canonical form of the EE is 
\begin{equation}
\eta _t+\eta \eta _z\ +\nabla ^2\eta _z+\epsilon \left( \eta _{zz}+\nabla
^4\eta \right) =0.  \label{e34}
\end{equation}
The control parameter in this equation is 
\begin{equation}
\epsilon =(1/3){\sqrt{4WR/5}.}  \label{(2)}
\end{equation}

Equations (1) and (2) have been {\it derived} directly from the fundamental
NS equations by using an iterative procedure which is a variation of the
so-called multiparameter perturbation approach (see e.g. \cite{f92}, \cite
{fi96}, and references therein; an earlier, more limited application of
multiple independent perturbation parameters appears e.g. in \cite{b66}). In
addition to its leading to the most general EE, another advantage of this
technique is that it yields the least restrictive conditions of theory
validity. For the present case, they require that the following two
dimensionless parameters be {\it independently} small: 
\begin{equation}
\alpha \equiv \sqrt{R/W}\ll 1~~~~~\mbox{and}~~~~~~~\beta \equiv \sqrt{R^3/W}%
\ll 1.  \label{(3)}
\end{equation}

From the linear stability theory, the third (third-order derivative) term of
Eq. (\ref{e34}) is purely dispersive, while all other linear terms are
dissipative. Different limiting cases of Eq. (\ref{e34}) reproduce some
known nonlinear equations, such as the 2D version of Korteweg-de Vries (KdV)
equation for $\epsilon \rightarrow 0$ and 2D version of Kuramoto-Sivashinsky
(KS) equation for $\epsilon \rightarrow \infty $ (see also \cite{ti89}). The
1D limit ($\partial _y=0$) of Eq. (2) is essentially the well-studied (see
e.g. references in \cite{one-d-theory}) equation first introduced by
Kawahara \cite{k83}.

To exhibit interesting spatial behavior, a system should be sufficiently
``large''. For the periodicity domain $0<y\leq 2\pi p$, $0<z\leq 2\pi q$ in
our simulations of Eq. (\ref{e34}), we chose $5\le p\le 16$ and $16\le q\le
80$. We used spatial grids of up to $32p\times 32q$ nodes, with the Fourier
pseudospectral method for spatial derivatives and with appropriate
dealiasing. Time marching was done (in the Fourier space) by using Adams-
Bashforth and/or Runge-Kutta methods. We checked the results by refining the
space grids and time steps; by verifying the volume conservation, $\int \eta
dydz=0$; etc. A typical simulation run took $\sim 10^6$ time-steps.

The initial values of $\eta $ were chosen independently at each node from
the interval [$-0.05,0.05$] with uniform probability distribution. Due to
the dissipativity of Eq. (\ref{e34}), the system evolves to an attractor,
and so essentially ``forgets'' the initial conditions. For large values of
the ``dissipativity'' parameter $\ \epsilon \gg 1$, as soon as the flow
approaches its asymptotic state, the observed film surface is 
irregular in space and time; no
spatial patterns are evident. The chaotic character of the
attractor is indicated
by the {\it positive} largest Liapunov exponent 
[which we found, similar to Deissler \cite{d89}, 
by numerically integrating, along
with Eq. (2), the linear equation that governs small disturbances
of the solution of Eq. (2)]. 
This is in accordance with the fact that in the limit $\epsilon 
\rightarrow \infty$,
Eq. (2) reduces to a 2D generalization of the Kuramoto-Sivashinsky equation,
whose solutions on extended spatial domains are known to exhibit 
chaotic attractors. Regarding the transient behavior on the way to the
attractor, our simulations of Eq. (2) with small $\epsilon$
corresponding to the parameter values of the experiment \cite{ls95}
have shown agreement \cite{fi96}
with their {\it transient} ``3D'' patterns and pattern
transitions, including checkerboard patterns, synchronous instabilities, and
solitary waves.

The main focus of the present communication is the presence of highly
nontrivial orderly patterns in time-asymptotic states for the {\it strongly
dispersive} cases, $\epsilon \ll 1$. Figure 1 shows snapshots of the film
surface at large times for three different sets of parameter values. (The
fact that by those times the systems have approached their asymptotic states
is clear, e.g., from the corresponding plots of the evolution of ``energy'' $%
\int \eta ^2dydz$ [see Fig. 2 corresponding to Fig. 1(a)]. We will speak of
such numerically identified time-asymptotic states as {\it attractors},
although one needs to be cautious here: it is known that such extended
systems may sometimes exhibit long transients. We find the largest Liapunov
exponent to be positive in this case as well, suggesting a {\it strange}
attractor.)

There are two subpatterns in Fig. 1(a): The V-shaped formation consisting of
13 large-amplitude bulges aligned into two straight lines moves as a whole
downstream with a certain velocity, and the small-amplitude lattice-like
subpattern of bumps moves uniformly as well, but in the opposite (in our
reference frame) direction. [Similar segregation of coherent structures into
two subpatterns of different amplitudes is also seen for the non-square,
large-aspect-ratio domains, Figs. 1(b) and 1(c).] This collision-course
movement is evident in the cross-sectional space-time portrait shown in Fig.
3. Even though each bump changes its shape in irregular manner, the bump
maintains its identity. In particular, the bumps do not seem to coalesce or
break up, and just weakly interact with one another. Also, the height of a
bulge {\it irregularly fluctuates}, by an amount which is approximately
equal to the amplitude of the bumps. 

As a bump runs into a bulge, the bulge's amplitude increases momentarily,
and then decreases again as a bump separates from the opposite side of the
bulge (see Fig. 3). These interactions, unlike the irreversible coalescences
of 1-D pulses--- discovered in \cite{kf94} for a highly nonlinear
dissipative equation---appear to be (almost) reversible, like the well-known
interactions of 1D KdV solitons.

We note that {\it bulge} formations similar to that of Fig. 1(a) were
discovered in \cite{ti89} for $\epsilon ^{-1}=25$ and $p=64/(2\pi )$ (they
postulated an equation of the form (2) based on an equation of the form (1)
derived in Ref. \cite{tk78} for a partial case of an inclined film; in fact,
that derivation was {\it not} valid for the {\it vertical}
film). But the authors
of \cite{ti89} seem to have overlooked (perhaps, because of inadequacy of
the graphics tools they used) the second, bump subpattern---and thus the
entire complex, dynamical character of the two-component order.

It is natural to inquire as to how the various quantities of the pattern
scale with $\epsilon .$ We varied $\epsilon ^{-1}$ between $25$ and $305$
for $p=q=16$. In one set of simulations, $\epsilon ^{-1}$ was gradually
decreased from $50$ in relatively small steps of $5$ (to allow the system to
``adiabatically'' adjust to the new parameter value), up to $\epsilon
^{-1}=25$---at which point the line formations of bulges break down. In
another set of simulations, $\epsilon ^{-1}$ was increased from $50$ in
steps of $10$ or $15$ up to $\epsilon ^{-1}=305$. In all cases, we find that
the characteristic width of the bulge as well as the bump is of the order of
($\sim$) $1$ independent of $\epsilon $. The amplitude of bulges is also
constant, $\sim 1 $, as are the velocity of bulges and that of bumps. 
Only the amplitude of
bumps changes; it scales like $\sim \epsilon $.

The V-shaped formation of bulges retains its form when $\epsilon $ is
changed from $1/30$ to $1/305$. However, the (absolute value of the) angle $%
\varphi $ of each bulge line with the streamwise axis decreases with $%
\epsilon $, probably approaching some asymptotic value in the limit $%
\epsilon \rightarrow 0$ (see Fig. 4; since there is no parameters remaining
in this limit, the asymptotic angle should be just $0$). It might be
possible to explain this dependence by a theory of pairwise interaction of
bulges through their (non-axisymmetric) ``tails'' (similar to the theory 
\cite{akpn} for 2D chemical-wave spirals).

When $\epsilon \ll 1$, the dissipative terms in Eq. (\ref{e34}) can be
treated as perturbations $\sim \epsilon $ of the 2D KdV equation 
\begin{equation}
\eta _t+\eta \eta _z+\nabla ^2\eta _z=0.  \label{e2dkdv}
\end{equation}
This equation does not seem to have any analytical solutions. However, by
transforming to a reference frame moving with a velocity $c>0$ [replace $%
\eta _t$ with ($-c\eta _z$) in the equation], Petviashvili and 
Yan'kov \cite
{py82} numerically obtained a stationary axially-symmetric solitary-wave
solution. By balancing the first term with the nonlinear term, $c\eta _z\sim
\eta \eta _z$, and the latter term with the dispersive term, the
characteristic amplitude and velocity of these solutions are found to be $%
\eta \sim c$ and $c\sim 1/L_s^2$ where $L_s$ is the characteristic
lengthscale, which is not uniquely determined by the KdV equation (\ref
{e2dkdv}). However, the two dissipative terms of Eq. (\ref{e34}) will change 
$L_s$ on a slow time scale, until they balance each other (this essential
role of the small dissipative terms was revealed in Ref. \cite{k83} for the
1D case). This selects the soliton of $L_s\sim 1$, which results in $c\sim 1$
and $\eta \sim 1$ as well, independent of $\epsilon $. These estimates are
clearly consistent with the numerical results for bulges reported above.

Motivated by the discovery of the second, small-amplitude subpattern, we
examined the possibility of a corresponding second travelling-wave solution.
If we transfer to the frame moving with a {\it negative} velocity $c=-a^2$,
where $a$ is a (real) constant, there are such solutions---with the
nonlinear term being as small as the dissipative ones. Indeed, the
leading-order equation then is $\nabla ^2\eta _z+a^2\eta _z=0$, which is the
well-known Helmholtz equation for $\eta _z$. There are solutions $\propto
\sin Jy\sin Kz$ ($J^2+K^2=a^2$). The balance between the (small) dissipative
terms again determines $K\sim c\sim 1$, and the balance of the dissipative
terms with the nonlinear term yields $\eta \sim \epsilon $. We see that
these lengthscale, amplitude and velocity (including its sign) agree with
those observed for the bumps in the numerical experiments as described
above. Note that our assumption of the {\it \ negative }velocity is
essential: with a positive velocity, one arrives at the {\it modified}
Helmholtz equation, which does not have any oscillating solutions. There are
only {\it exponential} solutions, which are unsuitable here. [We note that
the Helmholtz equation has axially-symmetric solutions as well, $\propto
J_0(ar)$ where $J_0$ is the Bessel function ($r$ is the radial coordinate).
This solution is only weakly localized: it decays at spatial infinity like a
power rather than exponentially. There is no such localized solutions in the
1D case, $\partial _y=0.$]

One would naturally like to find some known types of patterns to which those
reported here can be compared. There are several known cases (see e.g.  
Ref. \cite{lac} and references therein) of indefinitely long {\it coexistence}
of different types of patterns. 
However, in those cases each of the coexisting
patterns is confined to its own spatial region: its constituent structures
do not penetrate inside any ``alien'' pattern region. In contrast, we have
seen that the bumps constantly pass through the region of bulges. Another
possibility would be to look at the bulges and bumps as two traveling waves.
However, in contrast to usual cases, the bulge ``wave'' is confined to an
essentially 1D region, and there is a constant nonlinear interaction with
the wave of bumps.

Similar to Eq. (\ref{e34}), we have derived an equation for a film flowing
down a vertical {\it cylinder} (see Ref. \cite{fi96} and references
therein). In particular, one can see that if the (dimensionless) radius $b$
of the cylinder is not too small ($b\gg \beta ^{-1}$), the flow is well
approximated by the planar-film equation (\ref{e33}) [with periodic BC in the
azimuthal direction; we note that this also justifies our use of {\it %
spanwise-}periodic BCs in the numerical simulations. As to the streamwise
BC, we believe the solution becomes essentially insensitive to their
specific type in the limit of large aspect ratio $q/p$, as e.g. in Fig. 1.
It would be interesting to check this with spatial-evolution simulations,
such as those already conducted \cite{do91} for a different EE that
coincides with the non-dispersive limit of the above-mentioned EE \cite{fi96}].
One finds that with $h_0$ $\sim 1$ mm, the cylinder (dimensional) radius $%
\overline{b}\sim 1$ cm, and under parametric conditions $\alpha \ll 1,$ $%
\beta \ll 1$, and $\epsilon \ll 1$, for the waves (evolving as they
propagate from the entrance end of not-too-long a cylinder to its exit end)
to have enough time for approaching the attractor stage, the liquid should
be several hundred times as viscous as water. For example, it could be
glycerin with an admixture of water.

As a general conclusion, numerical 2D simulations of a realistic evolution
PDE signal that nonequilibrium dissipative systems can spontaneously form 
{\it non}periodic, but nevertheless highly ordered spatial patterns (of
compactly localized, soliton-like structures) which are of a remarkable 
complexity. In particular, the novel patterns consist of {\it subpatterns}%
---each of a different amplitude and each moving as a whole with its own
velocity, {\it penetrating} through one another. Thus, these subpatterns
coexist as the constituent components of the overall---microscopically{\it \
nonsteady} but macroscopically permanent---self-organized spatiotemporal
order characteristic of the system motion on such an attractor.

The particular dissipative-dispersive PDEs (1) and (2) have been
consistently derived from the full Navier- Stokes problem to provide a
controllably close approximation to the evolution of a liquid film flowing
down a vertical plane. The unconventional perturbative approach used in
this derivation has the advantage of yielding the least restrictive
conditions of the validity of the theory. To satisfy those validity
conditions for a possible (terrestrial) experiment designed to observe
patterns of the novel type on a film flowing down a vertical cylinder, the
film liquid should be much more viscous than water; fortunately, suitable
liquids are readily available.

We are grateful to 
I. Yakushin for technical
assistance. We have used computing facilities of the Alabama Supercomputer
Authority and NERSC of the Department of Energy. 
This work was partly supported
by DOE Grant  DE-FG05-90ER14100.

\begin{figure}[tbp]
\caption{Snapshots of the time-asymptotic film surface self-organized in
simulations of Eq. \ref{e34}, for three different cases (bulges move down
the page here; for convenience of presentation, different axes may have
different scales; in reality, all ``bulges'' and ``bumps'' have small slopes
and are nearly axisymmetric). (a) $p=q=16$, $\epsilon ^{-1}=50$, and $%
t=1.6\times 10^5$; (b) $(p,q)=(16,80)$, $\epsilon ^{-1}=30$, and $%
t=5.98\times 10^4$; (c) $(p,q)=(5,60)$, $\epsilon ^{-1}=25$, and $%
t=4.89\times 10^5$.}
\end{figure}
\begin{figure}[tbp]
\caption{ Evolution of the surface deviation ``energy'' $\int \eta ^2 dy dz$
from an initial small-amplitude ``white-noise'' surface to an attractor of
Eq. (2). The snapshot Fig. 1(a) was taken near the end of this run. Note
that the time unit here is $50$ times that of Eq. (2).}
\end{figure}
\begin{figure}[tbp]
\caption{Time-sequence of instantaneous surface profiles in a fixed vertical
cross-section normal to the film (for $p=q=16$, $\epsilon^{-1}=50$; the time
shown as $0$ is in fact $1.6\times 10^5$ counting from the start of the
run). In particular, it is evident that the (large-amplitude) bulges move in
one direction and (small-amplitude) bumps in the opposite direction.}
\end{figure}
\begin{figure}[tbp]
\caption{Angle $\varphi $ between (each) line of bulges and the streamwise
direction varies with $\epsilon $ ($p=q=16$).}
\end{figure}

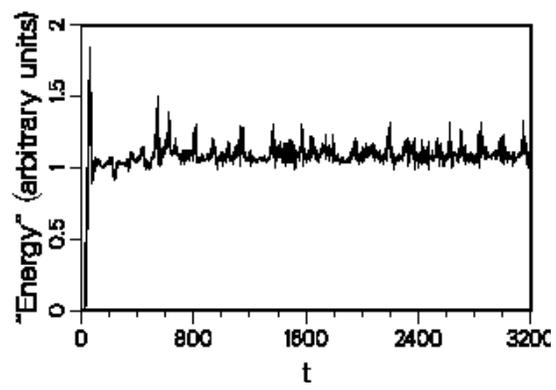
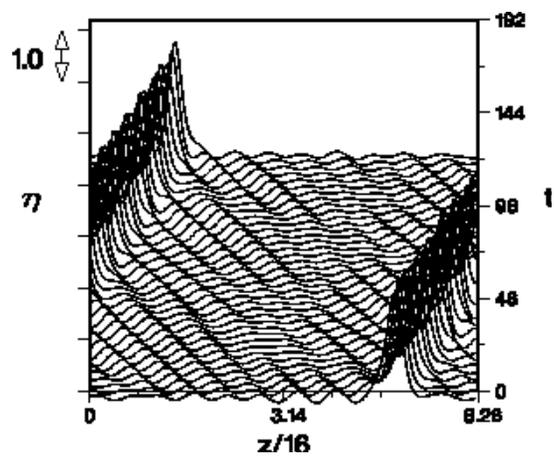
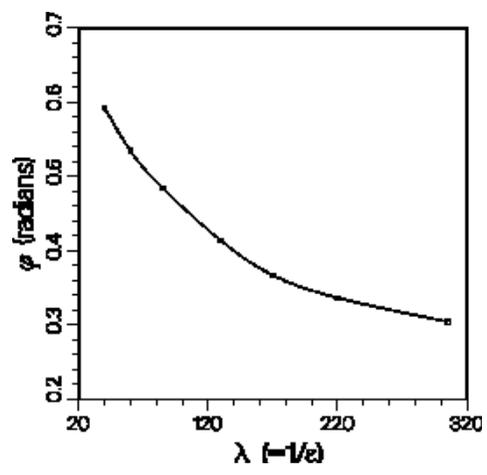
\end{document}